\documentclass[sigconf,10pt,nonacm]{acmart}
\AtBeginDocument{%
  }

\usepackage{color,soul}
\usepackage[font=small]{caption}
\usepackage{tcolorbox}

\usepackage{enumitem}
\usepackage{xspace}
\usepackage{tikz}
\usetikzlibrary{tikzmark}

\newcounter{mymarkzero}
\newcommand\startunderlinezero{\stepcounter{mymarkzero}\tikzmark{startmarkzero-\themymarkzero}}
\newcommand\stopunderlinezero{\tikzmark{stopmarkzero-\themymarkzero}}
\newcounter{mymarkone}
\newcommand\startunderlineone{\stepcounter{mymarkone}\tikzmark{startmarkone-\themymarkone}}
\newcommand\stopunderlineone{\tikzmark{stopmarkone-\themymarkone}}
\newcounter{mymarktwo}
\newcommand\startunderlinetwo{\stepcounter{mymarktwo}\tikzmark{startmarktwo-\themymarktwo}}
\newcommand\stopunderlinetwo{\tikzmark{stopmarktwo-\themymarktwo}}
\newcounter{mymarkthree}
\newcommand\startunderlinethree{\stepcounter{mymarkthree}\tikzmark{startmarkthree-\themymarkthree}}
\newcommand\stopunderlinethree{\tikzmark{stopmarkthree-\themymarkthree}}

\usepackage{fontawesome}
\usepackage{xcolor}
\definecolor{lightgray}{gray}{0.9}
\definecolor{lightblue}{rgb}{0.9,0.9,1}
\definecolor{blue_bg}{rgb}{0.85,0.85,1}
\definecolor{lightyellow}{rgb}{1,1,0.8}
\definecolor{lightpurple}{rgb}{1,0.85,1}
\definecolor{red}{rgb}{1,0,0}
\definecolor{darkgreen}{rgb}{0.4,0.7,0.3}

\newcommand{\prog}[1]{\textsf{#1}}
\newcommand{\code}[1]{\texttt{\small #1}}

\definecolor{mGreen}{rgb}{0,0.6,0}
\definecolor{mGray}{rgb}{0.5,0.5,0.5}
\definecolor{mPurple}{rgb}{0.58,0,0.82}
\definecolor{backgroundColour}{rgb}{0.95,0.95,0.92}

\newcommand{\remove}[1]{}

\newcommand{\paragraphb}[1]{\vspace{0.05in}\noindent{\bf #1}\hspace{0.05in}}

\newcommand{\Name}{\textsf{MappedTrace}\xspace}

\begin{document}

%%
%% The "title" command has an optional parameter,
%% allowing the author to define a "short title" to be used in page headers.
\title[\Name: Tracing Pointer Remotely with Compiler-generated Maps]
{\Name: Tracing Pointer Remotely\\ with Compiler-generated Maps}

%%
%% The "author" command and its associated commands are used to define
%% the authors and their affiliations.
%% Of note is the shared affiliation of the first two authors, and the
%% "authornote" and "authornotemark" commands
%% used to denote shared contribution to the research.

%\author{Submission \#12 Anonymous Author(s)}

\author{
    \textbf{Zhiyao Ma, Caihua Li, Lin Zhong} \\
    Department of Computer Science \\
    Yale University \\
    \texttt{\{zhiyao.ma, caihua.li, lin.zhong\}@yale.edu}
}

\renewcommand{\shortauthors}{Zhiyao Ma, Caihua Li, Lin Zhong}

%%
%% The abstract is a short summary of the work to be presented in the
%% article.
\begin{abstract}
  Existing precise pointer tracing methods introduce substantial runtime overhead to the program being traced and are applicable only at specific program execution points.
We propose \Name that leverages compiler-generated read-only maps to accurately identify all pointers in any given snapshot of a program's execution state.
The maps record the locations and types of pointers, allowing the tracer to precisely identify pointers without requiring the traced program to maintain bookkeeping data structures or poll at safe points, thereby reducing runtime overhead.
By running the tracer from a different address space or machine, \Name presents new opportunities to improve memory management techniques like memory leak detection and enables novel use cases such as infinite memory abstraction for resource-constrained environments.
\end{abstract}

\settopmatter{printfolios=true}

%%
%% This command processes the author and affiliation and title
%% information and builds the first part of the formatted document.
\maketitle

\section{Introduction}
\label{sec:introduction}

Pointer tracing plays a vital role in memory management and program analysis.
Existing approaches usually introduce substantial runtime overhead to the program being traced and render tracing feasible at only specific program states.
We propose an innovative approach to pointer tracing through an extensive use of compiler-generated maps, which we refer to as \Name.
The goal of \Name is to accurately identify the locations of all pointers given any program execution state, with the tracer potentially residing in a different address space or on a different machine than the program being traced.

\Name enables pointer tracing by relying on read-only maps generated by the \emph{compiler} which describe the locations and types of pointers.
To generate these maps, the compiler assumes a set of preconditions which can be enforced at compile time.
This method eliminates the need of maintaining bookkeeping data structures or polling at specific program points, thereby minimizing runtime overhead.

Leveraging the compiler-generated maps, the \emph{tracer} identifies pointers given a snapshot of program execution state and outputs a directed graph describing the point-to relationship.
The ability to run the tracer asynchronously from the traced program or on a different machine enables a broad range of use cases, such as extending memory for resource-constraint devices, detecting memory leaks with little runtime overhead, and enhancing dynamic taint analysis.

In the following sections, we first demonstrate \Name's utility through various use cases (\S\ref{sec:use-cases}), then review the background of pointer tracing (\S\ref{sec:background}), delve into the design of \Name (\S\ref{sec:design}), and finally discuss the associated challenges and limitations (\S\ref{sec:challenges}).
Through this exploration, we aim to provide a novel framework that not only advances the current state of pointer tracing but also opens new opportunities for efficient and secure software development.

\section{Use Cases}
\label{sec:use-cases}

\paragraphb{Infinite Memory for Microcontrollers}
Microcontrollers are resource constrained computers, typically having memory of a few hundreds of kilobytes.
Without a memory management unit (MMU), the embedded operating system (OS) and applications execute in a single physical address space.
\Name enables the opportunity to provide an infinite memory abstraction to applications on microcontrollers without compromising runtime performance in the common case.

The infinite memory abstraction works analogously to page swapping on systems with virtual memory.
Microcontrollers should still keep their working set in local memory for optimal runtime performance.
When local free memory is unable to satisfy an allocation request, the embedded OS kernel swaps selected allocated memory chunks to a remote server, then updates the pointers that point to the swapped chunks to sentinel values. 
When application code dereferences a pointer with a sentinel value, the execution traps to the kernel, so that the kernel can bring the memory chunk back to local and update the pointers accordingly.

\Name provides the mechanism to locate affected pointers after moving an allocated memory chunk.
The remote server stores the compiler-generated maps and runs the tracer.
To provide the snapshot states, the embedded OS kernel maintains a persistent network connection to the server, through which the server can query registers values in application threads or memory values in the address space.

\paragraphb{Memory Leak Detection}
Rust is an emerging programming language that supports automatic resource management without relying on a garbage collector.
However, programs may leak memory even when written in safe Rust~\cite{rust-box-leak}.
Applying \Name to Rust programs allows examination of memory leaks without compromising runtime performance.

Leak detection works by comparing the set of allocated memory chunks maintained by the allocator against the set of reachable chunks identified through \Name.
Leaked chunks are those in the allocated set but not in the reachable set.

To prevent disruption of program execution, leak detection can work on a duplicated snapshot instead of halting the program.
The snapshot can be a core dump image of the running process or be provided by the debugger attached to a forked subprocess.
The original process continues execution without being affected by the tracer.

\paragraphb{Dynamic Taint Analysis (DTA)}
DTA is a program analysis technique that tracks the propagation of tainted data during execution, which is widely used in security domains like vulnerability detection~\cite{schwartz2010all}.
A dynamic taint policy specifies how taints are introduced, propagated and checked.
Once a taint checker detects illegal use of tainted data, it may halt the execution and further analyze the propagation to identify the vulnerability.
In this case, \Name offers a graph reflecting all existing pointers precisely, which assists analysis of propagation with pointer-based indirect memory accesses.

\section{Background and Related Work}
\label{sec:background}

Pointer tracing aims to identify all pointers in a given snapshot of program execution state.
The tracing typically begins with a set of pointers identified in function call stacks and the static data regions, then transitively follows these known pointers to discover more in reachable objects.
This technique has been extensively employed by garbage collectors to free unreachable objects, using either a precise or imprecise variant.
Our proposed \Name aims to minimize runtime overhead and enable a broader range of use cases.

\paragraphb{Precise Pointer Tracing}
Precise pointer tracing \emph{accurately} identifies the locations of pointer-typed values, guided by the type information generated by compiler. Its tracing result is \emph{complete}, including all pointers reachable from the program code. The accuracy allows garbage collectors to avoid false positives on dead objects and in turn memory leaks, while the completeness enables moving live objects as in generational collectors~\cite{g1-gc,python-gc} which need to update all affected pointers.

Existing implementations of precise pointer tracing incur runtime overhead to enable identifying pointers in function call stacks.
One direction is to dynamically register live pointers in stacks at runtime~\cite{henderson2002accurate,rafkind2009precise}.
Another direction leverages compiler-generated information, called \emph{stack maps}~\cite{llvm-stack-maps}, to locate pointers in the stacks, but the maps are available only at a select few instruction addresses called \emph{safe points}.
Runtime overhead persists as typical implementations of safe points require spilling pointers in registers to the stack frames.
In multi-threaded programs, every thread must also poll at each safe point to check if it should suspend, which is necessary when pointer tracing has been initiated by another thread.

We propose \Name to reduce runtime overhead with greater assistance from the compiler.
At every instruction instead of only at safe points, there will be a map describing the locations of pointers in the stacks or registers.
This approach not only allows pointer tracing to start at an arbitrary program state, but also avoids the overhead of spilling and polling at safe points or registering pointers during execution.

\paragraphb{Imprecise Pointer Tracing}
Imprecise pointer tracing treats all values within a certain range as potential pointers.
Although this technique avoids the runtime overhead associated with precise pointer tracing, the inclusion of non-pointers in the tracing results limits its applicability to conservative garbage collection~\cite{boehm1988garbage}.
\Name opts for precise tracing to enable a broader range of use cases.

\section{Design}
\label{sec:design}

We propose \Name, a technique designed to accurately identify the locations of all pointers remotely given any execution state of a compiled program.
Being remote, the \emph{tracer} that identifies pointers may reside in a different address space or on a separate machine from the program being traced.
Being precise, the tracer reports only live pointers and ensures that no pointers are missed.

To minimize runtime overhead for the traced program, the tracer does not require any communication for bookkeeping during program execution.
Instead, the \emph{compiler} generates read-only maps that are stored on the tracer side, facilitating the discovery of pointers.
Programs to be traced must adhere to three precondition rules (\S\ref{subsubsec:design-rules}), which enable the compiler to generate these maps.
The compiler may enforce these rules at compile time (\S\ref{subsubsec:design-rule-enforce}).
Notably, we observe that the Rust compiler enforces a superset of rules that extends beyond the requirements of \Name.
Guided by the compiler-generated maps (\S\ref{subsubsec:design-maps}), the tracer operates on a snapshot of the program to identify existing pointers and generates a directed graph describing their relationships (\S\ref{subsubsec:design-tracing}).

\subsection{Precondition and Enforcement}

We first present the precondition rules that a program must adhere to in order to enable \Name, followed by a discussion on the opportunities to enforce these rules at compile time.

\subsubsection{Precondition Rules}
\label{subsubsec:design-rules}

\Name requires the program to adhere to three precondition rules, ensuring the compiler reports all potentially valid pointers to the tracer via the generated maps.
In cases where the tracer requires runtime information to determine a pointer's validity, these rules ensure the tracer can verify the pointer's validity by comparing its value with \code{null}.
Lastly, the rules require a \code{pass} operation to prevent race conditions related to pointer validity.

\paragraphb{Pointer Location (R1)}
To enable the compiler to identify the locations of pointers, only pointer-typed variables are permitted to store values intended for use as pointers.
The compiler maintains bookkeeping of locations only for pointer variables.
Storing pointer values in integer types causes the compiler to treat them as integer values, leading to missed pointers and, consequently, an incomplete memory view.

To avoid type ambiguity of the union locations, the variant read from a union must be the one stored most recently.
The compiler can then determine the active variant by adding a discriminant field which gets updated upon each store.

\paragraphb{Pointer Validity (R2)}
To prevent tracing through dangling pointers, pointers must always point to an object of the correct type or be \code{null}.
In other words, pointers must not dangle, even if they are not dereferenced during the dangling period.
This rule is essential as the compiler cannot always determine at compile time if a pointer is live at a given point of execution, in which case the compiler conservatively includes such pointer in the generated maps, and later the tracer determines the pointer's validity by testing its nullity.

Ignoring the \code{pass} keyword for now, the following C code snippet demonstrates an example where, after the conditional call to the \code{free} function, it is impossible to statically determine whether the pointer \code{p} dangles.

\vspace{-3ex}
\begin{figure}[h!]
    \centering
    \includegraphics[width=\linewidth]{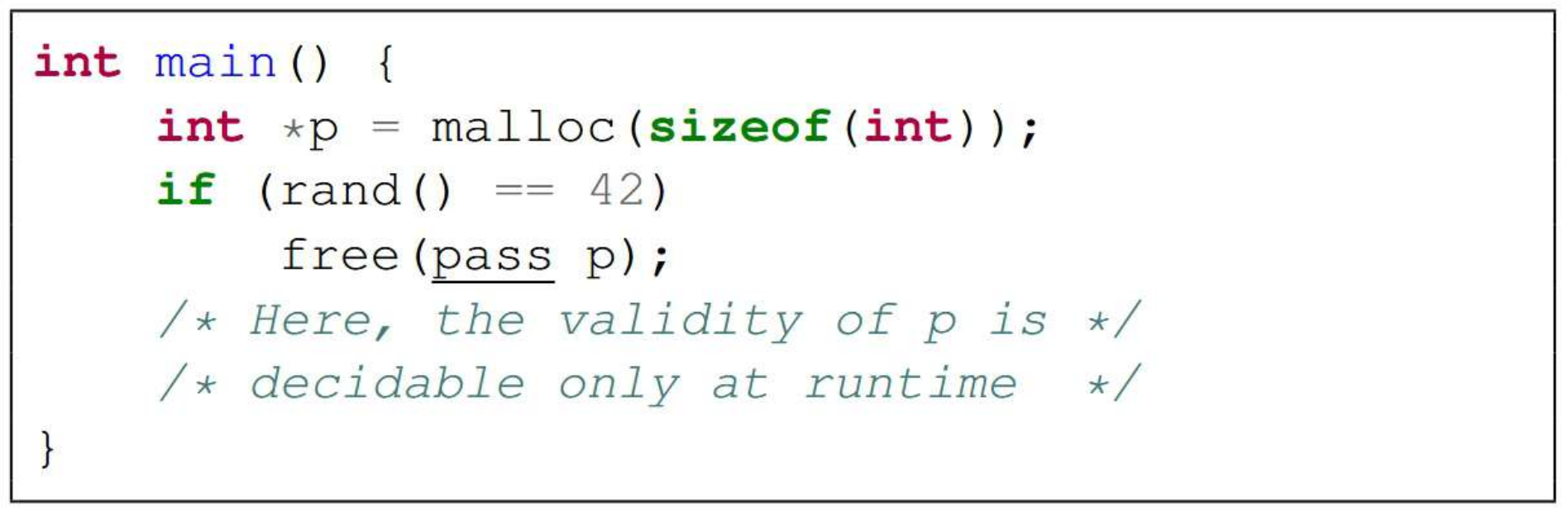} % Adjust width as needed
\end{figure}
\vspace{-3ex}

\begin{comment}
\vspace{-3pt}
\begin{minted}[escapeinside=++,frame=single,fontsize=\footnotesize]{c}
int main() {
    int *p = malloc(sizeof(int));
    if (rand() == 42)
        free(+\startunderlinezero+pass+\stopunderlinezero+ p);
    /* Here, the validity of p is */
    /* decidable only at runtime  */
}
\end{minted}
\vspace{-5ex}

% Draw the lines under the move keyword.
\tikz[overlay,remember picture]
 {\draw([yshift=-2pt]pic cs:startmarkzero-1)--([yshift=-2pt]pic cs:stopmarkzero-1);}
\end{comment}

The rule requires \code{p} to be \code{null} after being freed, but is violated after returning from \code{free} and before any following \code{null} assignment. Thus, the language must support a \code{pass} operation, as exemplified with the \code{pass} keyword above, which copies the current value of \code{p} and atomically sets it to \code{null}.

\paragraphb{Pointer Type (R3)}
Pointer types must match the types of the objects they point to.
The tracer later relies on this invariant to correctly identify the structure of the objects being pointed to, in order to discover pointer fields within \code{struct}s and transitively follow them.

\subsubsection{Compiler Enforcement}
\label{subsubsec:design-rule-enforce}

Since human programmers are prone to errors, we propose that the compiler enforce the precondition rules.
The compiler should prohibit casting to pointers to eliminate ambiguity in pointer values (R1 and R3) and restrict programming patterns to ensure pointer validity (R2).
We envision the compiler enforcing these rules for the majority of the source code while allowing programmers to introduce unchecked code blocks for greater flexibility if the compiler erroneously rejects the program due to incompleteness.
The Rust programming language has implemented a similar division between safe and unsafe code.
For the safe code, the Rust compiler enforces a superset of rules beyond those required by \Name.
Additionally, even without compiler enforcement, many well-known C programs are close to meeting \Name's preconditions.

\paragraphb{Disambiguate Pointer}
The compiler prohibits casting to pointers.
A value assigned to a pointer must meet one of the three conditions: it is \textit{(i)} copied from another pointer of the same type, or \textit{(ii)} returned from a function with the correct return type, or \textit{(iii)} the address of an object with the correct pointee type.
By restricting casting, only pointer-typed variables may hold values to be dereferenced, ensuring that their types match the objects they point to, thus satisfying the pointer location (R1) and pointer type (R3) rules.

To support memory-mapped I/O (MMIO), the compiler provides an intrinsic function that allows the definition of pointers to MMIO registers at fixed addresses.
Programmers should refrain from using this intrinsic function to define pointers that reference stack or heap memory regions.

Other compilers like the Rust compiler have already enforced such restrictions on casting.
It is impossible to cast other values into Rust references which are statically verified safe pointers.
Similarly, compilers that support precise garbage collections for C~\cite{henderson2002accurate,rafkind2009precise} also prohibit such casting.

Notably, even without compiler enforcement, many well-known C programs~\cite{libzip,gcc,bzip2,gzip,lame} contain no casting to pointers.
This is proven by the success of applying precise garbage collection to these programs without modifying their code~\cite{rafkind2009precise,banerjee2020sound}.
We believe restricting pointer casting does not significantly hinder the expressiveness of a language.

\paragraphb{Verify Pointer Validity}
The compiler restricts code patterns or requests annotations from programmers to determine pointer validity at compile time.
Note that the validity required by \Name is somewhat relaxed: Pointers must point to live objects of correct types only when they are not \code{null}.

The Rust compiler has demonstrated a way to ensure pointer validity by combining static lifetime analysis with a linear type system.
Lifetime analysis guarantees that Rust references are never \code{null} and always point to objects of correct types.
While the analysis is sound, it is incomplete, which means that the compiler may reject correct programs.
To enhance the language's expressiveness, Rust introduces smart pointer types such as \code{Box}, whose lifetime is tracked by the linear type system.
In cases where it is impossible to statically track the pointer's validity, the compiler inserts a boolean flag to ensure that the pointer is freed exactly once by the end of the scope, as shown below with the underline.
To satisfy the validity rule (R2), the compiler need only change to set the pointer to null instead of clearing the flag.

Even in the absence of strict compiler enforcement, many C programs adhere to the ISO C standard~\cite{c-standard}, which defines pointer rules closely aligned with \Name's requirements.
The standard requires C pointers to always point to an object of compatible types or one past the end in the case of array elements.
\Name merely prohibits the latter case.
Moreover, it is considered good practice to set freed pointers to \code{null} instead of dangling them.
Therefore, we believe that \Name's validity rule is feasible within common programming paradigms.

\vspace{-3ex}
\begin{figure}[h!]
    \centering
    \includegraphics[width=\linewidth]{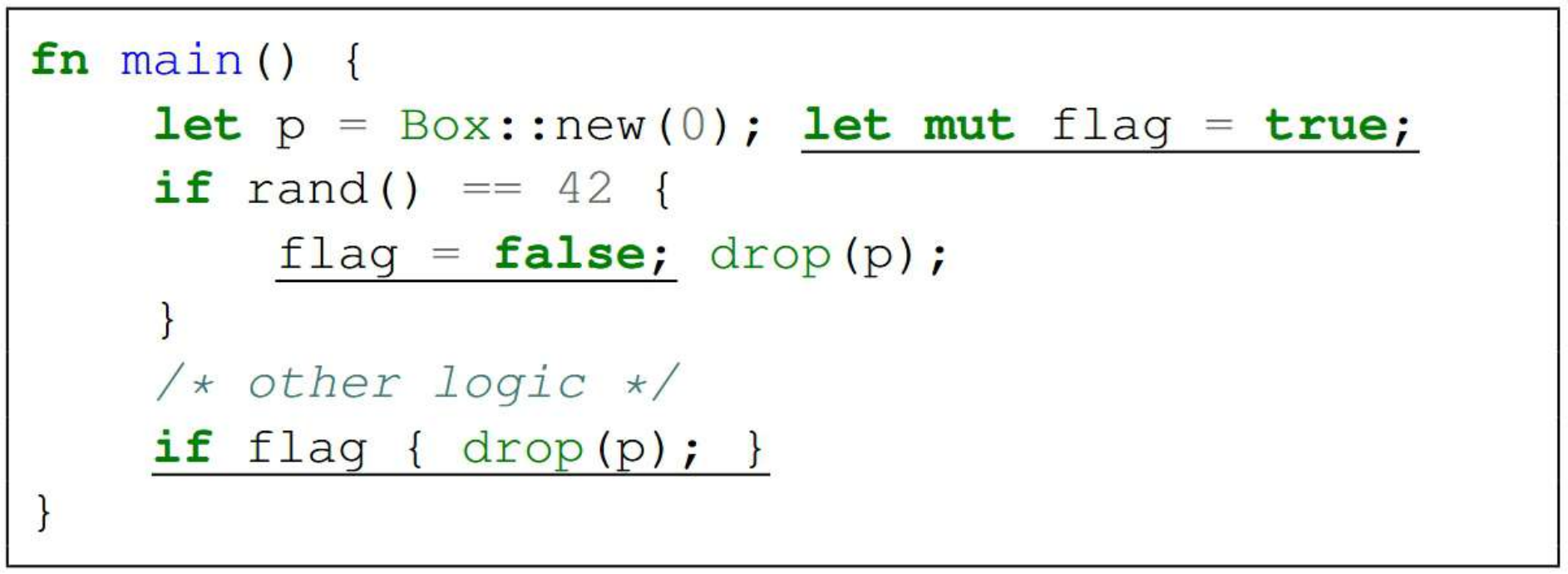} % Adjust width as needed
\end{figure}
\vspace{-3ex}

\begin{comment}
\vspace{-2ex}
% Draw the lines under the compiler inserted code.
\tikz[overlay,remember picture]
 {\draw([yshift=-2pt]pic cs:startmarkone-1)--([yshift=-2pt]pic cs:stopmarkone-1);}
\tikz[overlay,remember picture]
 {\draw([yshift=-2pt]pic cs:startmarktwo-1)--([yshift=-2pt]pic cs:stopmarktwo-1);}
\tikz[overlay,remember picture]
 {\draw([yshift=-2pt]pic cs:startmarkthree-1)--([yshift=-2pt]pic cs:stopmarkthree-1);}

\begin{minted}[escapeinside=++,frame=single,fontsize=\footnotesize]{rust}
fn main() {
    let p = Box::new(0); +\startunderlineone+let mut flag = true;+\stopunderlineone+
    if rand() == 42 {
        +\startunderlinetwo+flag = false;+\stopunderlinetwo+ drop(p);
    }
    /* other logic */
    +\startunderlinethree+if flag { drop(p); }+\stopunderlinethree+
}
\end{minted}
\end{comment}

\subsection{Pointer Tracing}

The tracer identifies all existing pointers given a snapshot of a program's runtime execution state facilitated by the compiler generated maps.
This snapshot can take various forms, such as a halted process attached to a debugger or a core dump image.
We first describe the read-only maps generated by the compiler and then explain how the tracer leverages these maps to locate pointers.

\subsubsection{Compiler-generated Maps}
\label{subsubsec:design-maps}

The compiler generates two categories of maps to facilitate the tracer: \emph{location maps}, which describe the locations of pointers within function stack frames, the static data region, or registers, and \emph{type maps}, which describe the offsets of pointers within a \code{struct}.
The compiler takes special care with union types to avoid type ambiguity and potential race conditions.
We also envision a series of optimizations to reduce the size of these maps.

\paragraphb{Generate Maps}
For every function it compiles, the compiler generates a location map that maintains the locations of pointers during the function's execution.
This map associates each program counter value within a function with a list of locations that store pointers.
A location is either a register, identified by its name, or a word in the stack frame, identified by its offset from the stack pointer.
The compiler possesses sufficient information to generate the location maps, as it schedules registers, allocates words in stack frames, and manages register spills.
Additionally, the compiler generates a location map for the static data region, reporting each pointer by its offset within the region.

The compiler also generates a type map that reports the offset of fields within a \code{struct} if they are pointers.
Primitive types are treated as \code{struct}s containing a single field.
A pointer of type \code{*T} is associated with the entry for type \code{T} in the type map.
This association applies to pointers reported by either the location map or the type map, subsequently allowing the tracer to transitively discover reachable pointers.

\paragraphb{Handle Unions}
To avoid type ambiguity in locations storing union-typed variables, the location maps and the type map also indicate the address of the discriminant field when reporting a pointer as a variant within a union.
This allows the tracer to confirm the active variant type by reading the discriminant field during tracing.

To satisfy the pointer validity rule (R2), code that stores a pointer in a union must strictly follow the sequence below: write zeros (\code{null}) to the data field, update the discriminant field to indicate the pointer variant, and finally write the pointer value to the data field.
This sequence of operations avoids the race condition where the tracer could observe the union as a pointer variant with a dangling pointer value.

\paragraphb{Reduce Map Size}
Although the location maps logically contain an entry for each program counter value, the compiler can employ several strategies to reduce their sizes, ensuring that the storage required for the maps is not excessive.
First, the compiler can exclude pointers from the location maps if they are known not to be live at compile time.
For example, pointers declared at the beginning of a function need not be included in the location maps until they are assigned a non-\code{null} value.
Additionally, pointers that are unconditionally assigned to \code{null} need not be included thereafter.
Second, the compiler can merge entries with identical lists of pointer locations.
Entries for adjacent program counters can share the same list if none of the instructions move pointers among registers or across the stack frame. Finally, the compiler can further compact the location maps by recording only the differences between adjacent entries.
Since instructions typically update one register or stack word at a time, adjacent entries are likely to share mostly identical information.

\subsubsection{Tracing Algorithm}
\label{subsubsec:design-tracing}

The tracer identifies all pointers within a snapshot of a program's runtime state.
It relies on the snapshot to provide the register values of all threads and the values in memory.
The tracer operates in three steps.
First, it unwinds the call stacks to determine the program counter values of active functions.
Next, it queries the location maps to find the initial set of pointers in registers, stack frames, and the static memory region.
Finally, using the type map, the tracer recursively identifies all other reachable pointers.
The output from the tracer is a directed graph that describes the existing pointers in the snapshot.

\paragraphb{Unwind the Stack}
The stack unwinding procedure identifies the program counter values for each active function in the call stack.
The tracer then uses these program counter values as keys to query the location maps.
Modern implementations of stack unwinding~\cite{llvm-libunwind,nognu-libunwind,boos2020theseus,ma2023panic} already rely on compiler-generated read-only data structures, usually called unwind tables, which are leveraged by the unwinder to retrieve the program counter values from registers or the stack.
Due to the similarity in structure, it is straightforward to incorporate the unwinder into the tracer.

\paragraphb{Find the Pointer}
To identify all pointers in the program snapshot, the tracer first determines an initial set of pointers and then follows them transitively to discover additional pointers.
The initial set consists of the pointers stored in registers, function call stacks, and the static data region.
For each non-\code{null} pointer, the tracer uses the type map to check if the object being pointed to contains pointers.
If so, the tracer recursively follows these pointers.
To avoid infinite loops, the tracer stops recursion upon reaching an already visited object.

When forming the initial set, the tracer should exclude functions belonging to the dynamic memory allocator to avoid identifying pointers that point to partially allocated or freed memory chunks.
A simple solution is to omit reporting any pointers in the location maps for functions performing memory management.
As a result, a pointer becomes visible to the tracer only after being returned from the allocation function or becomes invisible after being passed to the \code{free} function (via \code{pass}).

The tracer relies on the type map to provide additional information for objects whose structure may change dynamically, such as unions and dynamic arrays.
The type map indicates the location of discriminant fields of unions or the lengths of arrays, allowing the tracer to recursively locate any pointers within them.

\section{Challenges and Limitations}
\label{sec:challenges}

\paragraphb{Extending Compiler Infrastructure}
We expect a significant challenge in extending existing compiler infrastructure to support the generation of the maps discussed in \S\ref{subsubsec:design-maps}. Specifically, the compiler must book-keep additional metadata at each instruction for every register and stack frame location storing pointers. Also, when performing architectural specific optimizations after lowering the intermediate representation (IR) to target machine instructions, these optimization passes must be modified to keep the metadata updated. Existing support in popular compiler infrastructures like \prog{LLVM}~\cite{lattner2004llvm} and \prog{Cranelift}~\cite{cranelift} restrict the maintenance of such metadata to a few safe-points defined by the frontend language. However, \Name requires the metadata to be available at every target machine instruction.

\paragraphb{Supporting Assembly}
We foresee another challenge in supporting assembly code with \Name. System code may include assembly to configure hardware or maintain runtime environment that high level languages expect. Programmers writing assembly must manually convey the location of pointers to the compiler for generating the location maps. We envision a solution similar to call frame information (CFI) directives~\cite{cfi-directives}. These directives are written as pseudo-instructions, guiding the compiler to generate unwind tables for the assembly code by indicating the locations of preserved callee-saved register values. \Name requires a similar design of directives to indicate the locations of pointers.

\paragraphb{Modifying Source Code}
A limitation of applying \Name to C or unsafe Rust is that the source code needs modification to use the \code{pass} keyword in order to prevent dangling pointers. Rust reduces the burden of code rewriting since it discourages the use of unsafe code except absolutely necessary, and thus the majority of code written in safe Rust already meets the requirement of \Name as discussed in \S\ref{subsubsec:design-rule-enforce}.

\paragraphb{Distributing Libraries}
Another limitation of \Name is that libraries need to be distributed in source code or otherwise in binary built by an \Name-capable toolchain. The client of the library requires the presence of the location maps and the type map in the library to enable pointer tracing. We believe that using a designated compiler toolchain is not a burden for the library developers.

\paragraphb{Introducing Overheads}
Finally, \Name introduces small runtime overhead through setting pointers to null and updating the discriminant field of unions. \Name may also increase stack frame size by rendering some compiler optimizations infeasible as shown in the example below. The variable \code{i} and \code{p} are prevented from sharing the same register or memory location because otherwise such location will have an ambiguous type during the instructions at the comment line. Nevertheless, we believe these overheads are small and the traced program can gain performance by not spilling registers or polling at safe points.

\vspace{-3ex}
\begin{figure}[h!]
    \centering
    \includegraphics[width=\linewidth]{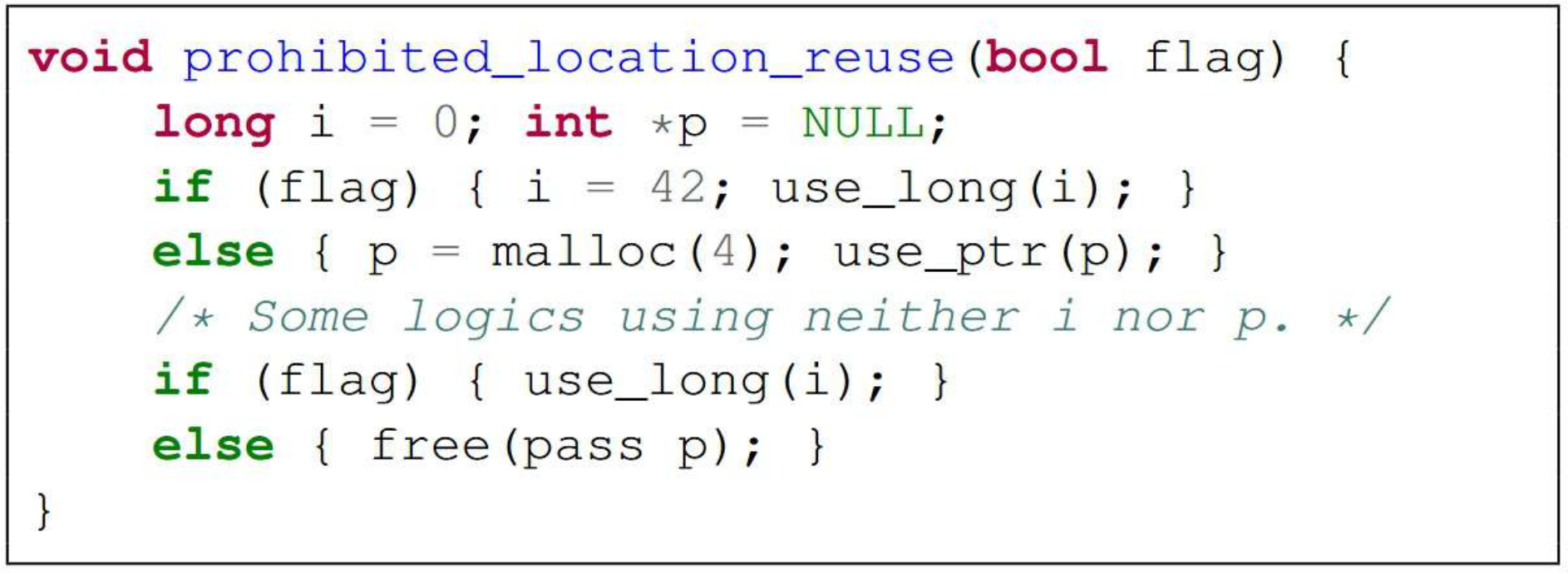} % Adjust width as needed
\end{figure}
\vspace{-3ex}

\begin{comment}
\begin{minted}[escapeinside=++,frame=single,fontsize=\footnotesize]{c}
void prohibited_location_reuse(bool flag) {
    long i = 0; int *p = NULL;
    if (flag) { i = 42; use_long(i); }
    else { p = malloc(4); use_ptr(p); }
    /* Some logics using neither i nor p. */
    if (flag) { use_long(i); }
    else { free(pass p); }
}
\end{minted}
\end{comment}

\section{Conclusion}

We propose \Name, a novel method to achieve precise pointer tracing with minimal runtime overhead.
Using compiler-generated maps, this approach addresses the limitations of existing methods, ensuring precision and efficiency.
The practical use cases of \Name are broad, such as providing infinite memory abstraction on microcontrollers, detecting memory leaks with little runtime overhead, and assisting dynamic taint analysis.
Implementing \Name requires extending current compiler infrastructures and potentially modifying source code to adhere to new precondition rules. 
Nevertheless, we believe the benefits of \Name outweigh the challenges in improving software efficiency and security, which signals a promising direction for future research and development.

%%
%% The acknowledgments section is defined using the "acks" environment
%% (and NOT an unnumbered section). This ensures the proper
%% identification of the section in the article metadata, and the
%% consistent spelling of the heading.
% \begin{acks}
% To Robert, for the bagels and explaining CMYK and color spaces.
% \end{acks}

%%
%% The next two lines define the bibliography style to be used, and
%% the bibliography file.
\bibliographystyle{ACM-Reference-Format}
\bibliography{sample-base,references}

%%
%% If your work has an appendix, this is the place to put it.
% \appendix

\end{document}